\documentstyle[11pt,psfig,paspconf]{article}

\begin{document} 
\title{A new catalogue of multiple galaxies in the Local Supercluster}
\author{D.~I.~Makarov, I.~D.~Karachentsev}
\affil{Special Astrophysical Observatory, Russia}

\begin{abstract}

  To reveal small galaxy groups in the Local Supercluster, a new 
approach is suggested which allows for individual properties of 
galaxies. The criterion is based on the assumption of closed motions 
of companions around the dominating group member within a zero 
velocity sphere.

  The criterion is applied to a sample of 6321 nearby galaxies with 
radial velocities $V_0\le3000$ km/s. The 3472 galaxies have been 
assigned to 839 groups that include 55\% of the considered sample. 
For the groups revealed by the new algorithm (with $k \geq 5$ 
members) the median velocity dispersion is 86 km/s, the median 
harmonic radius is 247 kpc, the median crossing time is 0.08(1/H), 
and the median virial-mass-to-light ratio is $56 M_\odot/L_\odot$.
\end{abstract}

\section{Grouping criterion}

Over the last years the number of galaxies with known radial 
velocities has grown greatly. This is true for distant volumes as 
well for the nearby volume of the Local Supercluster. For instance, 
the Nearby Galaxies Catalog by Tully (1988) contains  2367 galaxies, 
whereas last version of the LEDA Database (Paturel et al, 1996) 
collects about 6900 objects in the same volume of the Local 
Supercluster. The new observational data give us a ground to consider 
assignment of nearby groups again.

For selection of small galaxy groups we apply a new percolation 
algorithm basing on approach supposed by Karachentsev (1994). Unlike 
in earlier approaches, where individual properties of galaxies were 
used not enough, we consider pairwise interactions of galaxies and 
assume the full energy for each physical pair to be negative.

The condition of closed orbits can be expressed as the kinetic and 
gravitational energy ratio:

\begin{equation}
 \frac{\rm T}{\Omega}=\frac{V^2R}{2G\sum {\cal M}} < 1
\label{Crit1}
\end{equation}
where R and V are space linear separation and space velocity 
difference, $\sum{\cal M}$ is the total mass of the pair, and G is 
the gravity constant.  However, because of projection effects this 
condition itself does not allow to distinguish false pairs with small 
radial velocity difference.  Therefore, the condition (\ref{Crit1}) 
must be adopted with a limitation on the maximum galaxy separation in 
a pair.  Such a natural bound is the ``zero-velocity'' surface, which 
separates the collapsing volume against expanding space (Sandage, 
1986). In the case of  spherical symmetry expansion it can be 
expressed as:

\begin{equation}
 \frac{\pi^2R^3H^2}{8G\sum{\cal M}} < 1
\label{Crit2}
\end{equation}
where H is the Hubble parameter.

It should be noted that both the conditions are conservative with 
respect to projection factors, i.e. for each real bound pair they 
remain to be true. On the other hand, the group catalogue may still 
be polluted by optical pairs.

Our algorithm for group selection is a kind of percolation methods. 
On the first step it reveals pairs satisfying to the conditions 
(\ref{Crit1}) and (\ref{Crit2}).  At the second step all pairs with 
any common component link together into a group.  Finally, if a 
galaxy turns out to be a companion of several more massive galaxies, 
then we choose from these combinations the most massive attractor.  
Particularly, one group can form a subgroup inside more massive one.  
Therefore, our criterion combines the advantages of a 
``friends-of-friends'' companionship to a hierarchic ``dendrogram'' 
approach.

Each galaxy mass can be estimated from its luminosity or amplitude of 
its rotation curve. We use the first case, because it can be applied 
to all considered galaxies.  We account a mass-to-light relation for 
different galaxy types.  The ``total'' mass has been derived as

\begin{equation}
{\cal M} = \kappa {\cal M}_{25}
\end{equation}
where $1/\kappa$ is a fraction of mass within the standard galaxy 
radius.  According to numerous data on rotational curves of galaxies 
we assume as an average $\kappa\approx3$ (Hoffman et al., 1996; 
Broeils \& Rhee, 1997).

\section{Application of the criterion}

The described algorithm was applied to a sample of galaxies from the 
LEDA Database (Paturel et al, 1996), updated by the latest 
observations.

We selected galaxies with radial velocities less than 3000 km/s 
after correction for the Sun motion with respect to the Local Group 
centroid.  The region of ``Zone of Avoidance'' in the Milky Way  was 
excluded.

\begin{equation}
\left\{
\begin{array}{r@{ }l}
   V_0 < & 3000 \mbox{km/s,} \\
   |b| > & 10^\circ
\end{array}
\right.
\end{equation}

We did not remove the central part of Virgo cluster to check the 
method efficiency in crowded regions. Finally, 6321 galaxies were 
collected for analysis, which is 2.5 time their number in catalogue 
by Tully (1988).

Apparent magnitudes of galaxies were corrected for the Galactic 
extinction using new IRAS/DERBE map (Schlegel et al, 1998).  All 
other photometric corrections were made following the LEDA manner 
(Paturel et al, 1996).  To correct heliocentric radial velocities we 
used the apex parameters from Karachentsev \& Makarov (1996).  The 
Hubble constant of 70 km/s/Mpc was adopted.

\section{Results}

The criterion allows to identify 839 galaxy groups of different 
multiplicity.  At total these groups contain 3472 galaxies, i.e. 55 
percent of the considered galaxy sample. For different parameters of 
the groups their median values and quartiles are presented in the 
table.

{
\renewcommand{\arraystretch}{1.5}
\begin{table}[h]
\caption{The median parameters of the groups.}
\centerline{  
\begin{tabular}{rrcrlr}
\hline
\hline
\multicolumn{1}{c}{  k  }&
\multicolumn{1}{c}{  N  }&
\multicolumn{1}{c}{$M_v/\sum L$}&
\multicolumn{1}{c}{$\sigma_V$}&
\multicolumn{1}{c}{$R_h$}&
\multicolumn{1}{c}{$\tau_L H_0$}\\
\hline
 2 & 424 & $18_{-14}^{+39}$ &  $24_{-13}^{+22}$ & $170_{-114}^{+185}$ & $0.21_{-0.15}^{+0.45}$ \\ 
 3 & 158 & $32_{-22}^{+34}$ &  $41_{-18}^{+19}$ & $191_{-88}^{+157} $ & $0.15_{-0.09}^{+0.16}$ \\ 
 4 &  72 & $28_{-13}^{+39}$ &  $52_{-18}^{+30}$ & $187_{-77}^{+135} $ & $0.10_{-0.05}^{+0.10}$ \\ 
 5 &  47 & $42_{-24}^{+39}$ &  $57_{-14}^{+26}$ & $230_{-99}^{+92}  $ & $0.10_{-0.04}^{+0.12}$ \\ 
 6 &  27 & $50_{-31}^{+51}$ &  $87_{-32}^{+29}$ & $176_{-70}^{+93}  $ & $0.06_{-0.02}^{+0.06}$ \\ 
 7 &  21 & $52_{-12}^{+71}$ &  $96_{-24}^{+38}$ & $189_{-45}^{+150} $ & $0.06_{-0.01}^{+0.03}$ \\ 
\hline
$\ge5$ & 185 & $56_{-28}^{+40}$ & $ 86_{-28}^{+38}$ & $247_{-87}^{+89} $ & $0.08_{-0.03}^{+0.06}$ \\ 
$\ge8$ &  90 & $60_{-28}^{+38}$ & $102_{-33}^{+26}$ & $284_{-101}^{+83}$ & $0.08_{-0.03}^{+0.05}$ \\ 
$\ge20$&  11 & $80_{-39}^{+42}$ & $159_{-74}^{+23}$ & $298_{-10}^{+223}$ & $0.08_{-0.02}^{+0.03}$ \\ 
\hline                                                                           
all& 839 & $29_{-20}^{+41}$ &  $40_{-20}^{+35}$ & $194_{-106}^{+148}$ & $0.14_{-0.08}^{+0.22}$ \\ 
\hline
\hline
\end{tabular}
}
\end{table}
}

\begin{figure}
\centerline{  \psfig{figure=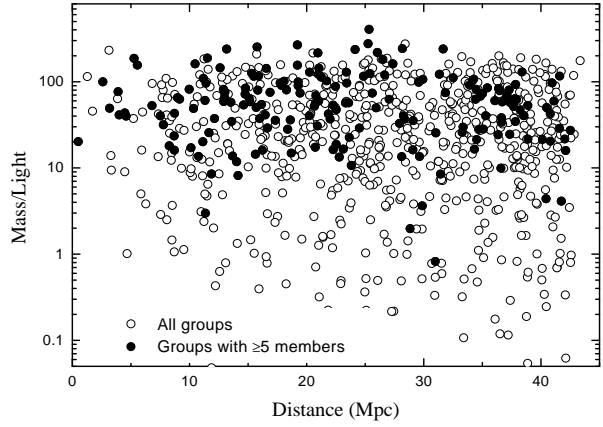,width=10cm}  }
\caption{Mass-to-light ratio versus distance to groups.}
\label{figMLD}
\end{figure}

In fact there is no correlation between velocity dispersion and 
distance to the groups as well between their virial-mass-to-light 
ratio and group distances (see fig.~\ref{figMLD}).  Therefore, we 
conclude that our clustering algorithm does not introduce 
considerable biases to the dynamical parameters of groups with their 
distance.

\begin{figure}
\centerline{
\begin{tabular}{p{6.5cm}p{6.5cm}}
\centerline{\psfig{figure=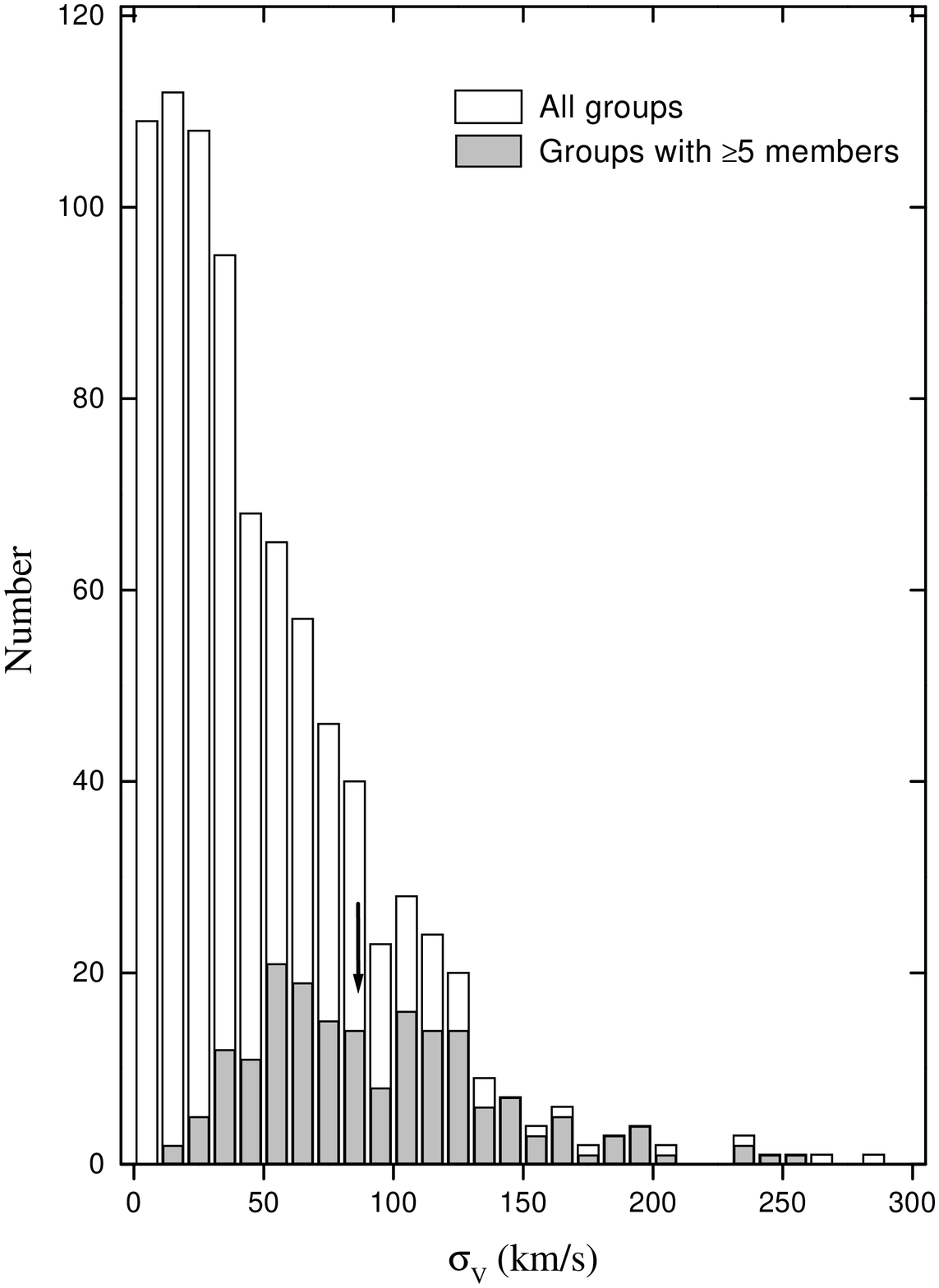,width=5cm}} &
\centerline{\psfig{figure=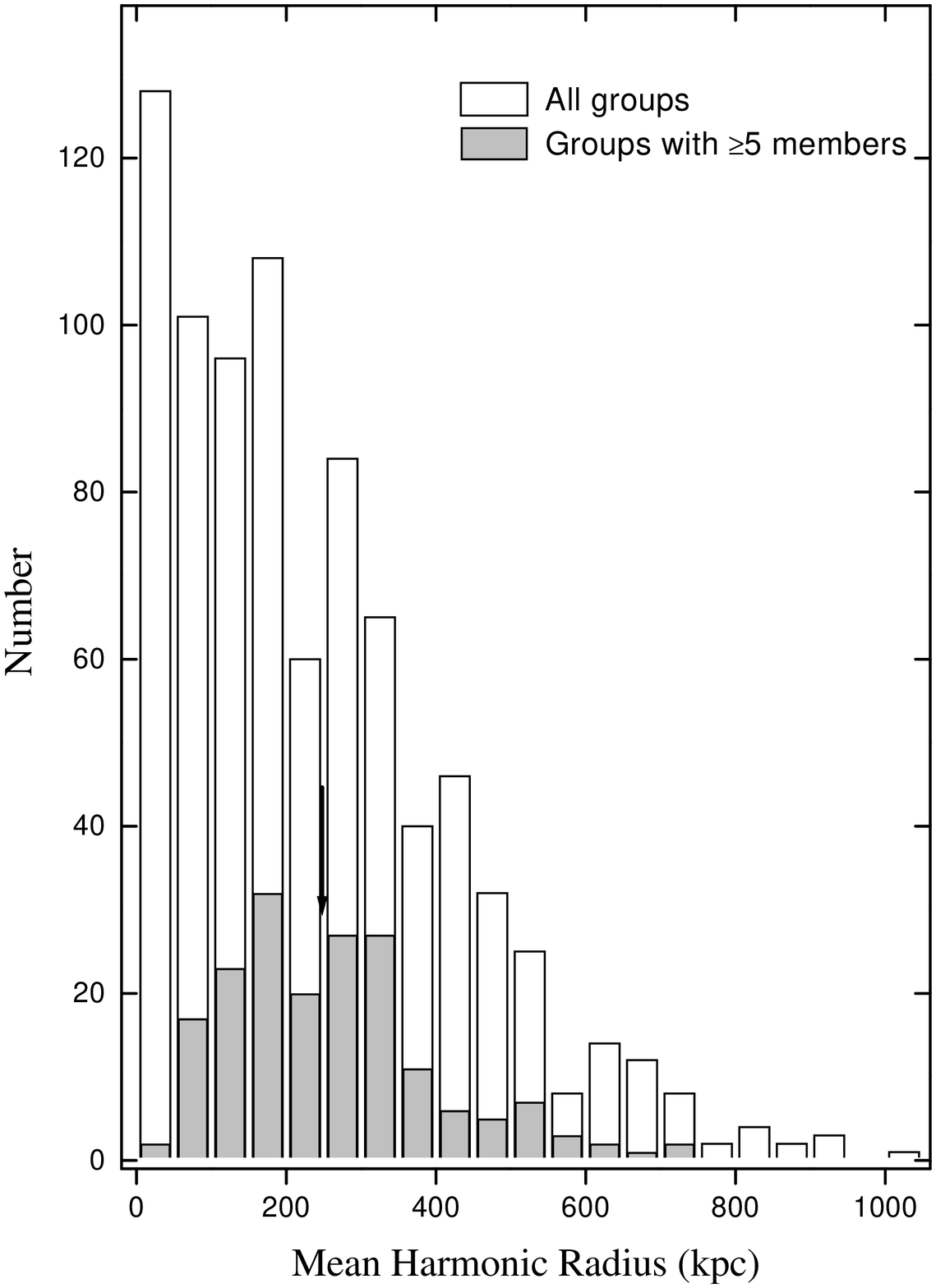,width=5cm}} \\
\caption{Histogram of velocity dispersion. The arrow indicates the 
  median value of 86 km/s for groups with $\ge5$ members.} 
  \label{figVsd}&
\caption{Mean harmonic radius. The arrow point out the 
  median value of $R_h=247$~kpc for groups with $\ge5$ members.}
  \label{figHR}
\end{tabular}
}
\end{figure}

\begin{figure}
\centerline{
\begin{tabular}{p{6.5cm}p{6.5cm}}
\centerline{  \psfig{figure=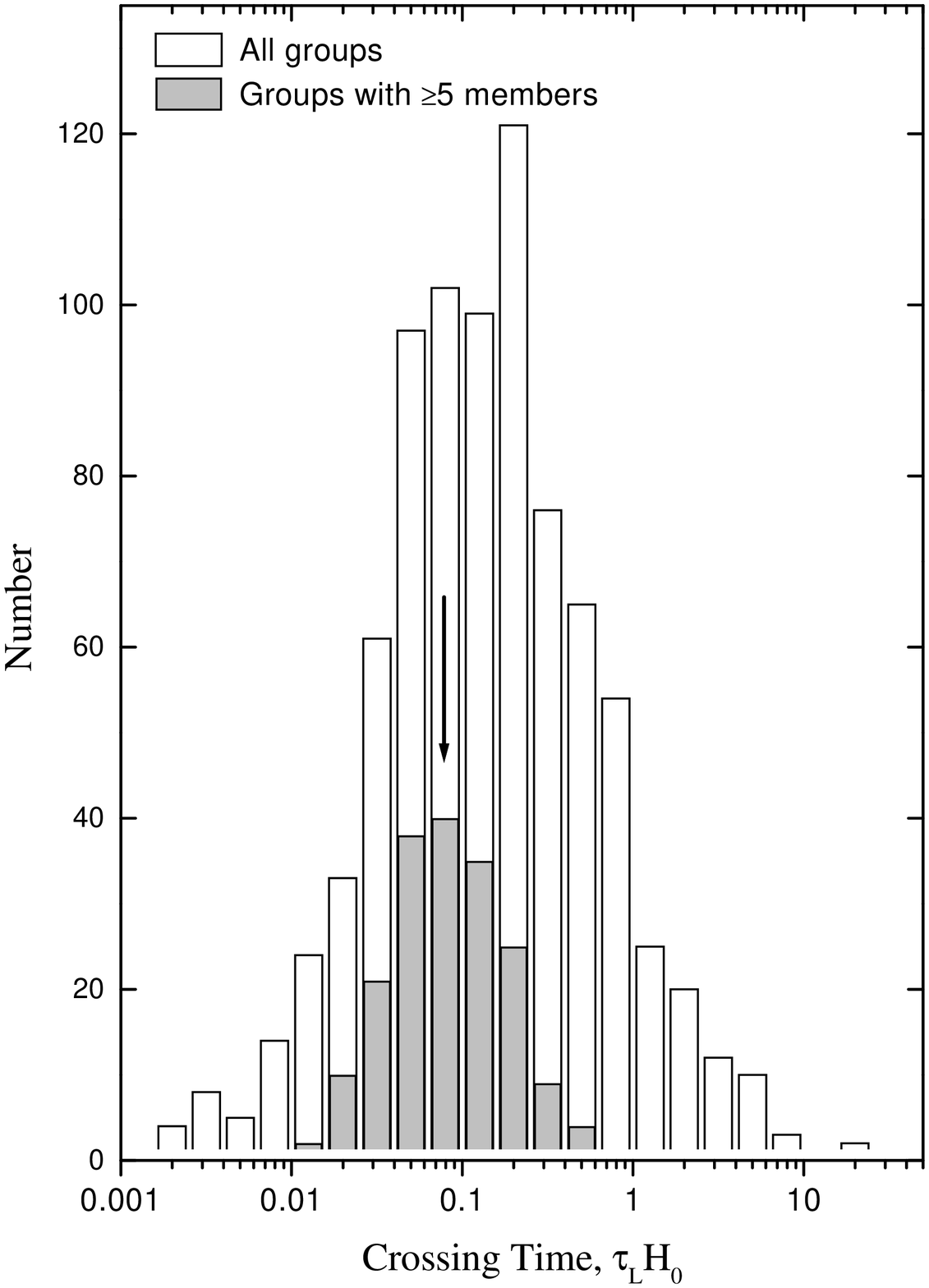,width=5cm}  } &
\centerline{  \psfig{figure=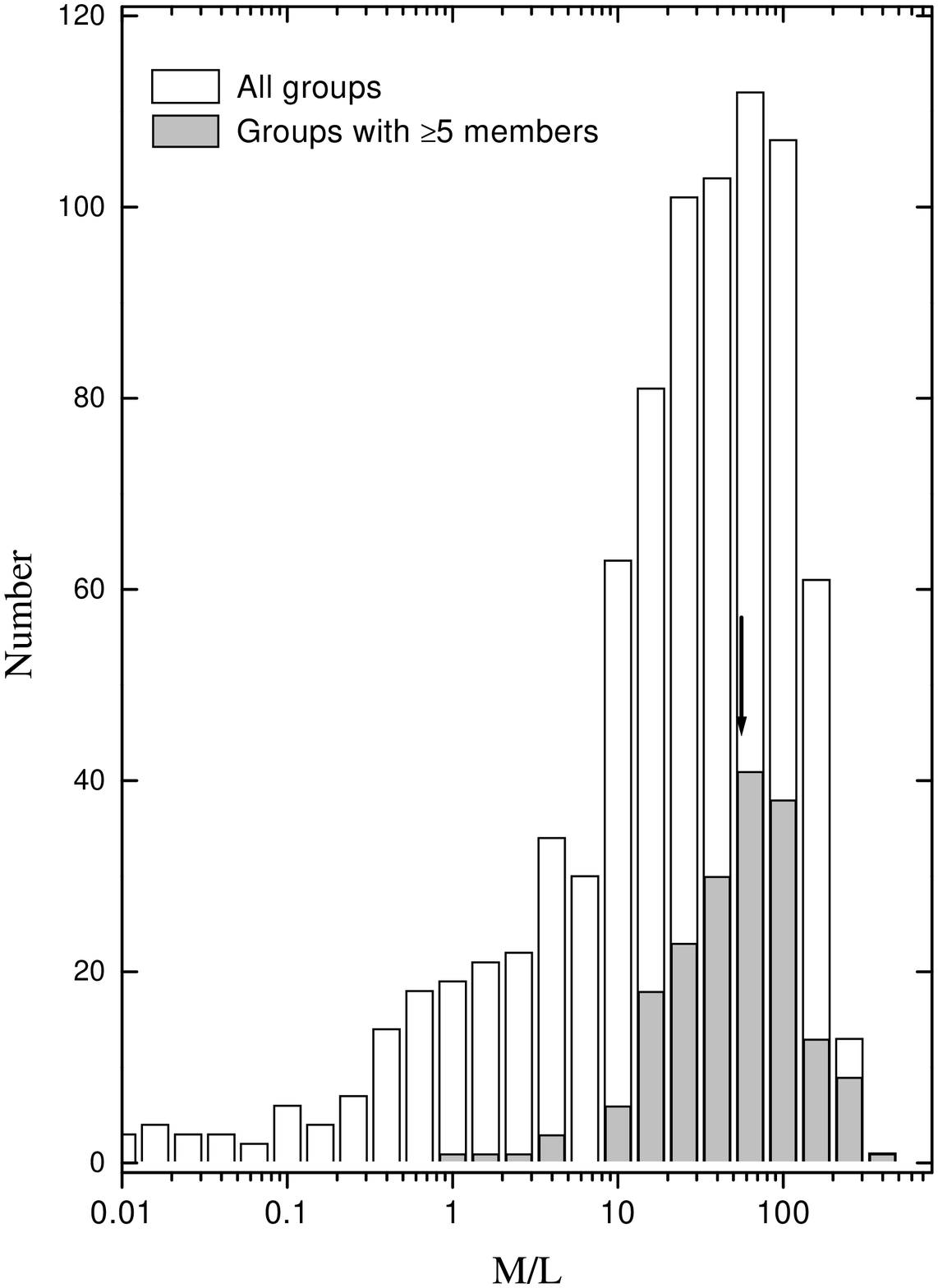,width=5cm}  } \\
\caption{The histogram of crossing time. The median value for the
  ``rich'' groups with $\ge5$ members of $\tau_L H_0=0.08$ is indicated 
  by the arrow.}
  \label{figCT} &
\caption{Mass-to-light ratio. The median value of 86~km/s
  for the large groups with $\ge5$ members is indicated by the arrow.}
  \label{figML}
\end{tabular}
}
\end{figure}

The median velocity dispersion in groups increases with group 
membership.  For groups with $k \geq 5$ it is 86 km/s only, being 
considerably lower than in the Geller \& Huchra (1983) catalog.  The 
maximum value of radial velocity dispersion, 287 km/s, is comparable 
with the rotational velocity amplitude typical for giant galaxies.  
The distribution of groups by velocity dispersion is presented in 
fig.~\ref{figVsd}.

The median harmonic radius (see fig.~\ref{figHR}) for galaxy 
systems equals to 247~kpc.

The median crossing time for our groups is only $0.08H$, that 
indicates to the virialized state of many galaxy group 
(fig.~\ref{figCT}).

The virial-mass-to-luminosity ratio grows with number of galaxies in 
group, and for rich systems with $k \geq 5$ its median is $56 
M_\odot/L_\odot$. The derived quantity indicates to existence of 
moderate amount of dark matter in the groups. The histogram of 
mass-to-light ratios is presented in fig.~\ref{figML}.

\begin{figure}
\centerline{  \psfig{figure=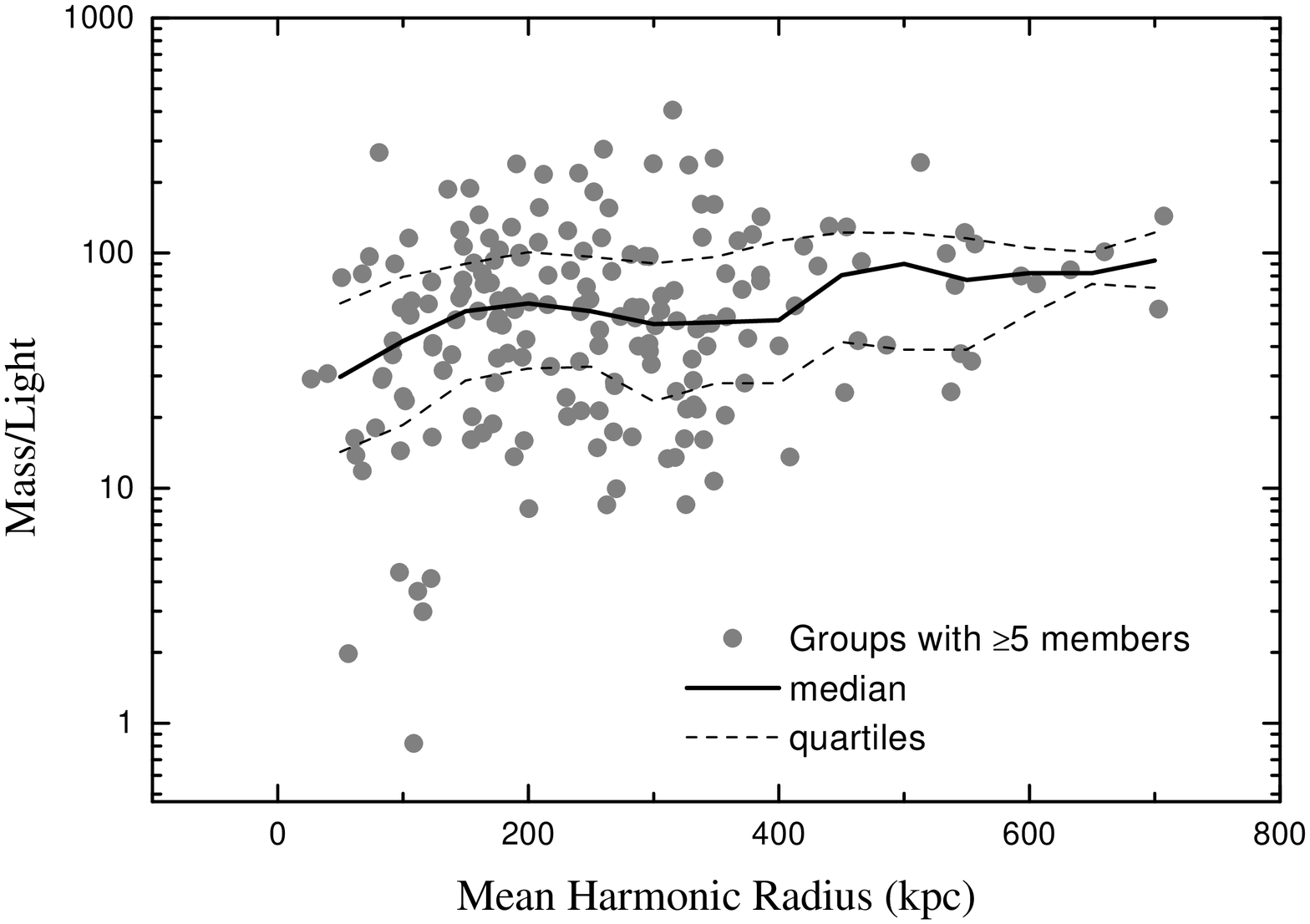,width=10cm}  }
\caption{Mass-to-light ratios as a function of harmonic radius of group.}
\label{figMLHR}
\end{figure}

As is seen in fig. \ref{figMLHR}, there is a tendency
of increasing of M/L ratio with harmonic radius of the groups.

Apparently, the proposed algorithm seems to be applied to low dense 
regions like the Local Group. But, in fact, it allows to 
distinguish some groups in Virgo cluster.  Two groups with population 
of about 80 members each were selected around the giant galaxies 
NGC~4486 and NGC~4472. These probable subclusters in the Virgo have 
been also noted by other authors.  However, one should be careful 
with the parameters of groups selected in the central part of 
rich clusters, because this method needs more physical ground to be 
applied to regions of high overdensity.

\section{Conclusions}

\begin{itemize}
\item We propose the algorithm for group selection, which accounts 
individual galaxy properties in their pairwise interaction and bases 
on a linkage of galaxies with negative mutual full energy.

\item For the assumed value of $\kappa=3$ about 55 percent of the 
galaxies have been grouped.

\item The dynamical parameters of the groups show no dependency on 
their distance.

\item In the rich groups the velocity dispersion is about $70\div80$ 
km/s, which differs considerably from the quantity obtained by Geller 
\& Huchra (1983).  The maximum velocity dispersion in the groups 
reaches 287~km/s.

\item The median crossing time is about 0.08 of the Hubble time that 
points to virialized state of many groups.

\item The median virial-mass-to-luminosity ratio is about 56, that 
shows a presence of moderate amount of dark matter in small galaxy 
groups.

\item The algorithm intend for small group selection, but probably it 
can be also applied to distinguish substructures in galaxy clusters.
\end{itemize}


\begin{references}
 
\reference Broeils A.\ H., \& Rhee M.-H., 1997, \aap, 324, 877.

\reference Geller M.\ J., Huchra J.\ P., 1983, \apjs, 52, 61

\reference Hoffman G.\ L., Salpeter E.\ E., Farhat B., Roos T.,
 Williams H., \& Helou G., 1996, \apjs, 105, 269

\reference Karachentsev I., 1994, A\&Ap Transaction, 6, 1

\reference Karachentsev I.\ D., Makarov D.\ I., 1996, \aj, 111, 794 

\reference Paturel G., Bottinelli L., Di Nella H. et al., 
  1996, Catalogue of Principal Galaxies: 
  PGC-ROM 1996, Saint-Genis Laval, Observatoire de Lyon

\reference Sandage A., 1986, \apj, 307, 1

\reference Schlegel D.\ J., Finkbeiner D.\ P., Davis M., 1998, \apj, 500, 525
 
\reference Tully R.\ B., 1988, Nearby Galaxies Catalog, Cambridge, 
  Cambridge Univ.\ Press
\end{references}
\end{document}